\documentclass[mypaper,7pt,twoside]{CoAst}
\usepackage{epsf,graphicx,fancyhdr}
\usepackage{natbib}
\renewcommand{\chaptermark}[1]{\markboth{#1}{}}

\def\kms {km~s$^{-1}$}

\newcommand{\kopf}{\small\itshape Comm. in Asteroseismology\\ Vol. number, publication date (will be inserted in the production process)}
\newcommand{\Authors}[1]{\begin{center}\normalsize\bf\sf #1 \end{center}}

\renewcommand{\author}[1]{\begin{center}\normalsize\bf\sf #1 \end{center}}
\newcommand{\Address}[1]{\begin{center}\small\sf #1 \end{center}}

\renewenvironment{abstract}{\section*{Abstract}\normalsize\sf}{}

\newcommand{\chapterDSSN}[2]{\chapter[\sf\normalsize #1\\ \footnotesize \hspace*{5mm}by #2 \sf\normalsize][]{#1\\}\rhead[\fancyplain{}{\sf\footnotesize \center{#1}}]{\fancyplain{}{\sffamily\thepage}}\lhead[\fancyplain{\kopf}{\sffamily\thepage}]{\fancyplain{\kopf}{\sf\footnotesize \center{#2}}}}

\renewcommand{\chaptername}{}
\newcommand{\acknowledgments}[1]{\vspace*{5mm}\noindent\begin{bf}Acknowledgments. \end{bf} #1}

\usepackage[T1]{fontenc}

\def\Oi {{\Omega_{\mathrm{i}}}}

\def\kms {{\mathrm{km}\,\mathrm{s}^{-1}}}
\def\msol {{\mathrm{M}_\odot}}

\def\ratiorot {{\Pi_{1/0}\,(\Omega)}}
\def\ratiorotdeg {{\Pi_{1/0}^{\mathrm{d}}\,(\Omega)}}

\def\cesam {{\sc cesam}}
\def\filou {{\sc filou}}

\begin{document}
\sf

\chapterDSSN{Analysis of $\ratiorot$ period ratios in the presence of near degeneracy}
{J.C. Su\'arez, R. Garrido, A. Moya}

\Authors{J.~C. Su\'arez$^{1,2}$, R. Garridol$^{1}$, A. Moya$^1$} 
\Address{$^1$ Instituto de Astrof\'{\i}sica de Andaluc\'{\i}a (CSIC), CP3004, Granada, Spain
         $^2$ LESIA, Observatoire de Paris-Meudon, UMR8109, France}

\noindent
\begin{abstract}
      In the present work we provide the preliminary results obtained when
      analysing the rotational Petersen diagrams when including the effects
      of near degeneracy. 
      We found that near degeneracy affects significantly the fundamental-to-first
      overtone period ratios, $\ratiorot$, showing wriggles in the Petersen diagrams. 
      Analysis of such wriggles
      reveals that they are mainly caused by the avoided-crossing phenomenon. 
      The size of wriggles seems to increase with the rotational velocity
      and could, in certain cases, invalidate any accurate mass and/or 
      metallicity determinations. Nevertheless, deep analysis of near degeneracy
      effects may allow us to obtain additional information on the mode
      identification of the radial modes and their corresponding coupled
      pairs, which would allow us to constrain the modelling.
\end{abstract}


\section{Introduction\label{sec:intro}}

In a previous work \citep[][hereafter, SGG06]{Sua06pdrot}, we showed the importance 
of taking the rotation effects into account (even for relatively slow rotating stars, 
as double-mode pulsators) especially when accurate metallicity and/or mass determinations 
are required. In that paper, 1O period ratios were calculated for different 
rotational velocities (RPD) and metallicities and then compared with standard 
non-rotating Petersen diagrams (PD). The difference in period 
ratios was shown to increase with the rotational velocity for a given metallicity. 
Differences in the period ratios were found to be equivalent to differences in the
metallicity.

The present work intends to complete SGG06's work by including the effects of near 
degeneracy in the computations of oscillations. Mode identification is essential for 
the correct use of PD. As shown by \citet{Soufi98} and \citet{Sua06rotcel}, 
near degeneracy effects cannot be neglected for asteroseismic
studies of slowly-to-moderately rotating stars. Near degeneracy affects
the small separations since it occurs for close modes (under certain 
selection rules, $\Delta\ell=0,\pm2$, and $\Delta\,m=0$). However, such an effect is 
far from being trivial and deserves special attention.

We examine in detail the effect of 
rotation on mass and metallicity diagnostics based on Petersen diagrams, focusing
on the influence of the near degeneracy effects on the period ratios
of radial modes. Indeed, important effects are expected to 
be found when near degeneracy is taken into account \citep{Alosha03}. In that 
work, very large and non-regular perturbations of such ratios were expected to occur.
We found such perturbations under the form of wriggles in the RPD. 
However such wriggles seem to be regular. A short discussion on the possible
origin of these wriggles is here provided. 

\section{The $\ratiorotdeg$ period ratios\label{sec:impact_Z}}

In order to study the effect of near degeneracy on RPD, we construct 
tracks of asteroseismic models for different mass, metallicity and
rotational velocity. Equilibrium models are computed with the evolutionary
code \cesam\ \citep{Morel97} for which a first-order effect of
rotation is taken into account in equilibrium equations. Uniform rotation and 
global conservation of the total angular momentum is assumed. Such models 
are the so-called 'pseudo-rotating' models \citep[see][]{Soufi98, Sua06rotcel}. 
Although the non-spherical components of the centrifugal acceleration are not 
considered, they are included as a perturbation in the oscillation frequencies 
computation. Computation of the oscillation spectra is carried out using the oscillation
code \filou\ \citep{filou,SuaThesis} which is based on a perturbative analysis 
and provides adiabatic oscillations, corrected for the effects of rotation up 
to the second order (centrifugal and Coriolis forces), including
near degeneracy effects.
\begin{figure*}
 \begin{center}
   \includegraphics[width=9cm]{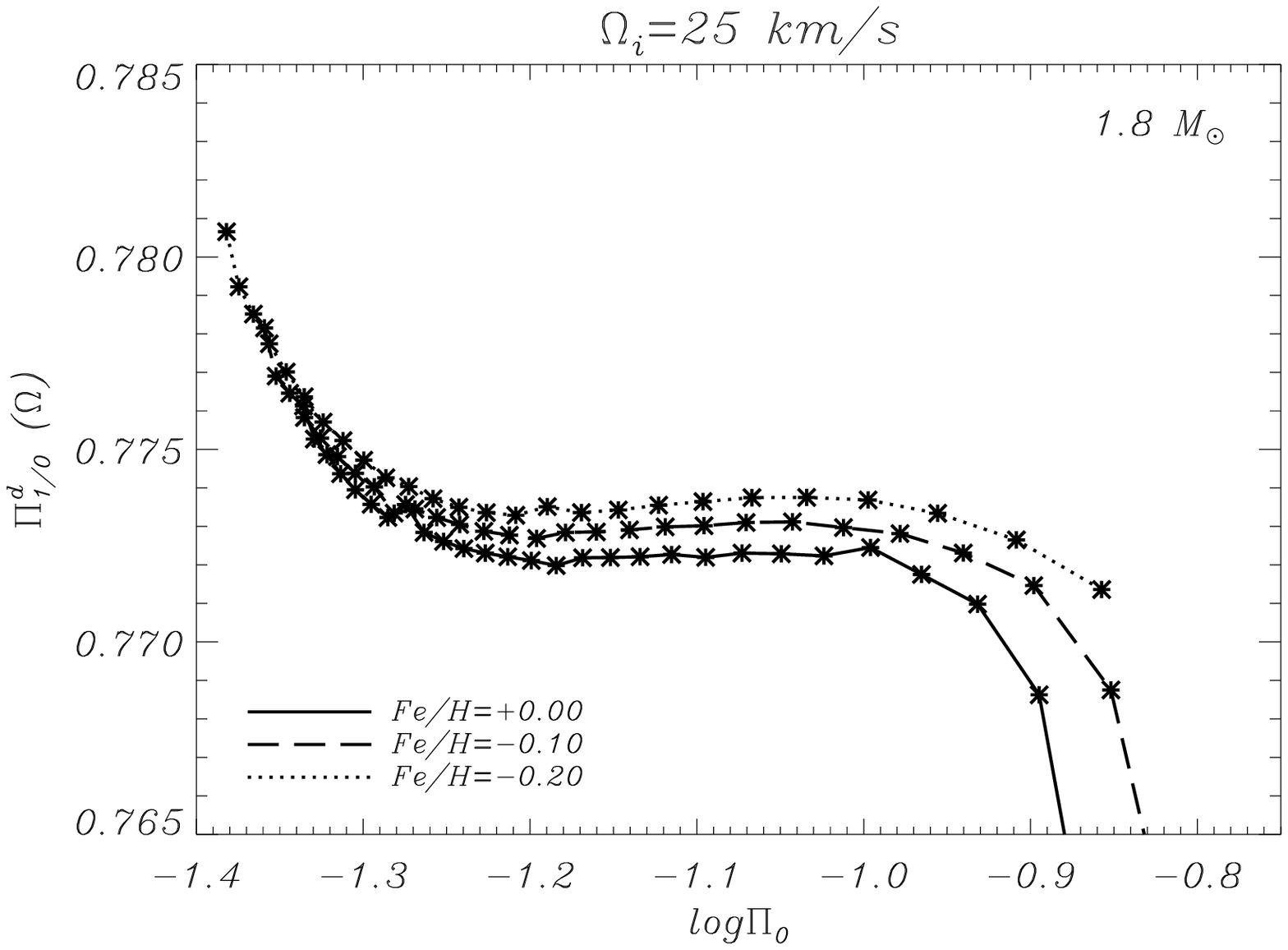}\hspace{-0.40cm}  
   \includegraphics[width=9cm]{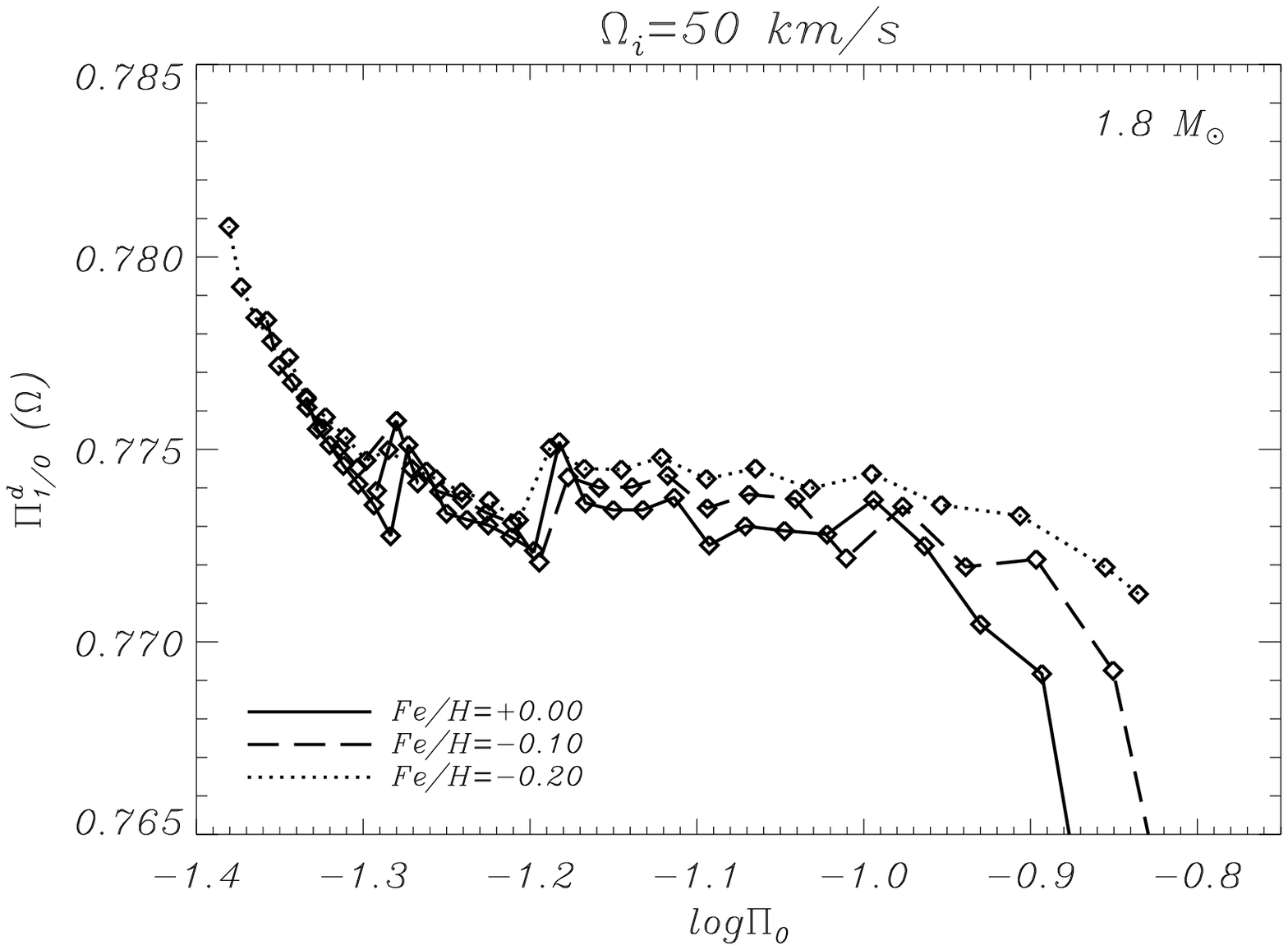}\hspace{-0.40cm} 
   \caption{Theoretical RPD showing the $\ratiorotdeg$ period ratios for a set
	    of evolutionary $1.8,\msol$ tracks obtained for different
	    metallicities. Tracks for two initial rotational velocities are 
	    considered: 25 and $50\,\kms$ (from top to bottom). For convenience, the 
	    following symbols are used: asterisks, representing models evolved with $\Oi=25\,\kms$;
	    diamonds and those evolved with $\Oi=50\,\kms$.}
   \label{fig:rot_PD}
 \end{center} 
\end{figure*}

In order to determine how near degeneracy affects mass and metallicity determinations
using RPD, we selected several evolutionary tracks with different metallicities
and two different initial rotational velocities $\Oi=25, 50$, for a
fixed mass of $1.8\,\msol$. For each model we then computed the corresponding
$\ratiorotdeg$ period ratio, i.e., the 1O period ratios including near degeneracy
effects. 

In Fig.~\ref{fig:rot_PD} we show RPD displaying
such period ratios, from top bottom, for tracks computed for $\Oi=25$ to $50\,\kms$ 
respectively. As expected, near degeneracy does not modify the general behaviour of 
the 1O period ratios with the metallicity, i.e., $\ratiorotdeg$ increases when 
increasing rotational velocities. This is equivalent to
decrease the metallicity in standard PD (see SGG06). However, the presence of 
wriggles may inverse this situation in the regions where the curves cross each other.
Wriggles are found larger for increasing rotational velocities and they become significant
(in the context of RPD) degrading substantially the accuracy of period ratios $\ratiorotdeg$,
which can reach up $3\cdot10^{-3}$. In terms of metallicity, this implies uncertainties reaching 
up to 0.50\,dex (for the largest rotational velocity considered) which is critical for
Pop.~I HADS. This new scenario would invalidate, a priori, the PD as diagnostic diagrams. 
However, wriggles in period ratios do not seem to be located randomly in PD. 
They depend on the frequency evolution of the quadrupole modes coupled
with the radial ones which are mainly dominated by the avoided-crossing
phenomenon, and they seem nearly independent of the rotational velocity and  
metallicity. If these results are confirmed, it would be possible to provide
some clues for the mode identification of the fundamental radial mode, the first overtone, 
and their corresponding quadrupole coupled modes, only using \emph{white} light 
photometry. 
We will provide a complete study of the effect of near degeneracy on RPD, for 
different metallicities, masses and rotational velocities in a forthcoming
paper \citep{Sua07pdrotII}, {\bf in which we show that wriggles in period ratios
may imply differences about $10^{-2}$ (for rotational velocities around $50\,\kms$)
when comparing with non-rotating PD}. 
In that work we also examine certain properties of the near degeneracy
effects, namely the mode contamination i.e., the weight of the original individual 
spherical harmonics describing the oscillation mode in the resulting coupled mode; 
and the coupling strength, i.e., the effect of near degeneracy on the oscillation
frequencies. Analysis of these properties seem to be a promising procedure, not 
only to retrieve the usefulness of PD, but also to provide additional information
for the complete mode identification, the rotational velocity and the
inclination angle of the star.

\acknowledgments{JCS acknowledges support by the Instituto de Astrof\'{\i}sica 
de Andaluc\'{\i}a (CSIC) by an I3P
contract financed by the European Social Fund and from the Spanish
"Plan Nacional del Espacio" under project ESP2004-03855-C03-01.}

\bibliographystyle{aa}
\bibliography{/home/jcsuarez/Boulot/Latex/Util/References/ref-generale}

\end{document}